\pdfoutput=1

\documentclass[12pt]{article}
\usepackage{graphicx}

\newcommand\pubdate{\today}

\def\Title#1{\begin{center} {\Large #1 } \end{center}}
\def\Author#1{\begin{center}{ \sc #1} \end{center}}
\def\Address#1{\begin{center}{ \it #1} \end{center}}

\newcommand\pubblock{\rightline{\begin{tabular}{l}  \\ 
         \pubdate  \end{tabular}}}
\newenvironment{Abstract}{\begin{quotation}  }{\end{quotation}}
\newenvironment{Presented}{\begin{quotation} \begin{center} 
             PRESENTED AT\end{center}\bigskip 
      \begin{center}\begin{large}}{\end{large}\end{center} \end{quotation}}

\begin{document}
\begin{titlepage}
 \pubblock
\vfill
\Title{Prospects of transverse $\Lambda$ and 
$\bar{\Lambda}$ polarization measurements at LHCb}
\vfill
\Author{Cynthia Nu\~nez}
\Address{University of Michigan \\ on behalf of the LHCb Collaboration}
\vfill
\begin{Abstract}
Transverse $\Lambda$ polarization observed over four decades ago contradicted expectations from early leading-order perturbative QCD calculations. Measurements of $\Lambda$ polarization from unpolarized $pp$ and $p$A collisions have been previously observed to increase as a function of $x_F$ and $p_T$ up to a few GeV range and approximately independent of beam energy. Recent studies have linked polarization to the process of hadronization, which describes how particular hadrons are formed from scattered quarks and gluons. The high energy of the LHC and the coverage and precision measurement possibilities from LHCb forward geometry are ideal for studying hyperon polarization as a function of both $p_T$ and $x_F$. This contribution presents the status and prospects of hyperon polarization measurements in $pp$, $p$Pb, Pb$p$, and fixed-target $p$A collisions at LHCb.
\end{Abstract}
\vfill
\begin{Presented}
DIS2023: XXX International Workshop on Deep-Inelastic Scattering and
Related Subjects, \\
Michigan State University, USA, 27-31 March 2023 \\
     \includegraphics[width=9cm]{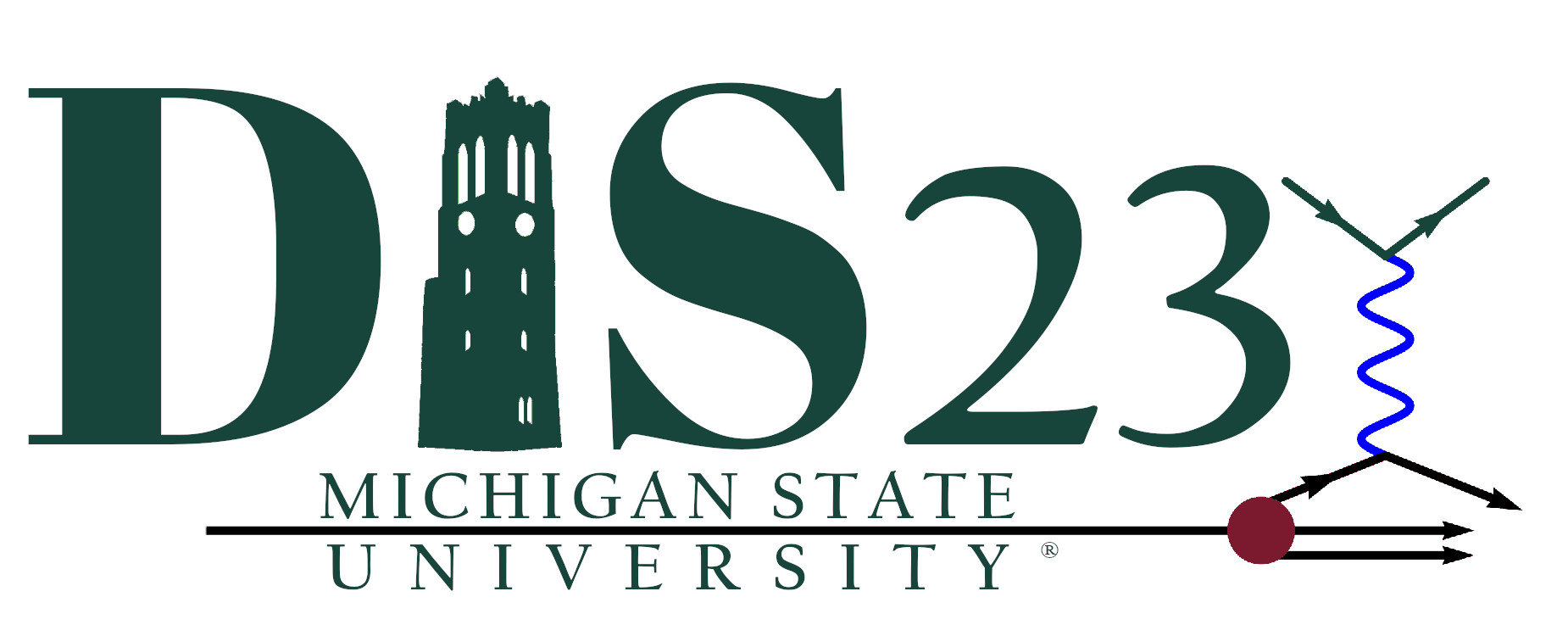}
\end{Presented}
\vfill
\end{titlepage}

\section{Motivation}
Transverse $\Lambda$ (uds) polarization was confirmed in 1976 in unpolarized proton-beryllium collisions, and was observed to be increasing up to $20\%$ for more forward $\Lambda$ production \cite{1976}. At the time, leading order perturbative QCD calculations predict very small polarization values \cite{kpr}. The polarization was also observed in various proton-nucleon ($p$A) and proton-proton ($pp$) experiments, observed to be negative, independent of the beam energy, and increases with increasing $|x_{F}|$ and $p_{T}$ up to measured values of a few GeV range. In 2018, $\Lambda$ and $\bar{\Lambda}$ polarization was observed in $e^{+}e^{-}$ collisions \cite{belle}, where since there is no initial state hadron there must be a hadronization effect. 

Non-zero spontaneous polarization for other hyperons has also been measured \cite{spin1996}. A similar polarization trend as the $\Lambda$ has been observed for $\Xi^0$ (uss) and $\Xi^-$ (dss), and with opposite sign for $\Sigma^+$ (uus) and $\Sigma^-$ (dds). The $\Sigma^0$ (uds) hyperon has the same polarization as $\Sigma^+$ and $\Sigma^-$, and opposite polarization than $\Lambda$ even though they have the same valence quark composition. The polarization has also been studied when the detected hyperons do not have valence quarks in common with the proton beam, and while zero polarization was observed for  $\Omega^-$ (sss) and $\bar{\Lambda}$ ($\bar{u}\bar{d}\bar{s}$), non-zero polarization was found for $\bar{\Xi}^-$ ($\bar{d}\bar{s}\bar{s}$) and $\bar{\Sigma}^+$ ($\bar{u}\bar{s}\bar{s}$). The spontaneous polarization for some hyperons but not others remains a question. 

The phenomenological approach to characterize the $\Lambda$ polarization has focused on looking at polarizing transverse-momentum dependent (TMD) fragmentation functions (FF) and higher twist multiparton correlators \cite{twist}. The polarizing TMD FF, $D_{1T}^{\perp \Lambda/q}(z, p_{\perp}^2)$, is a leading twist FF which describes the probability of an unpolarized quark hadronizing into a transversely polarized hyperon \cite{pff}. Higher-twist effects involving non-perturbative inputs aid in sensitivity to spin-momentum correlations in hadron structure and formation, and the twist-3 term is the next observable in the matrix element squared and refers to the interference term between scattering or hadronizing off one parton versus scattering off or hadronizing of two. For hyperon polarization, the polarization comes from either including higher order twist-3 collinear multi-parton correlation matrix elements or by convolution of a twist-2 TMD PDF with a twist-2 TMD FF.

\section{$\Lambda$ Polarization Measurement}
\begin{figure}[!h]
    \centering
    \includegraphics[width = 0.7\textwidth]{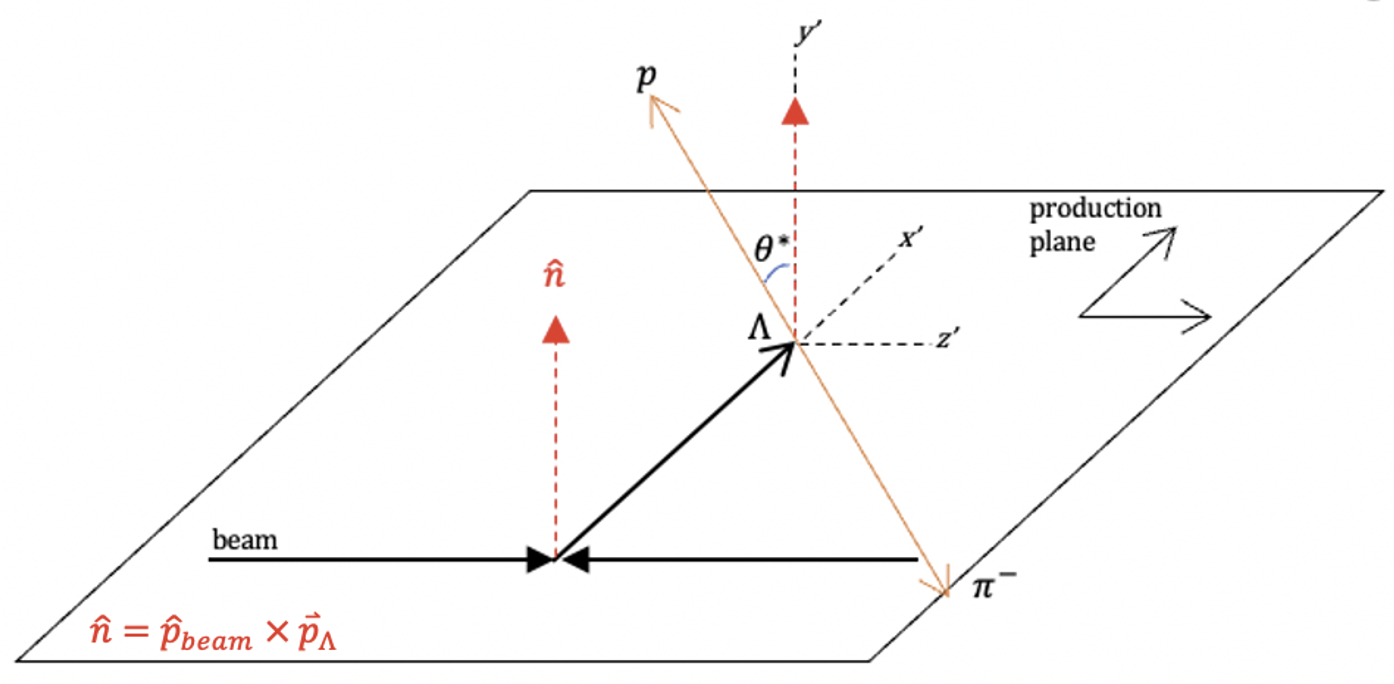}
    \caption{Diagram defining the polarization angle for the $\Lambda\rightarrow p\pi^-$ decay}
    \label{fig:geo}
\end{figure}

The $\Lambda$ polarization measurement uses the parity-violating weak decay nature of hyperons, where the polarization of the mother particle is revealed by analyzing the angular distribution of its decay products \cite{elliot}. For the $\Lambda\rightarrow p\pi^-$ decay, the transverse polarization is measured in the direction normal to the $\Lambda$ hyperon and beam momentum: 
$\hat{n} = \hat{p}_{beam} \times \vec{p_\Lambda}$. In the rest frame of the $\Lambda$ hyperon, the angle $\theta^*$ is defined between the decay proton momentum and the polarization direction, $\hat{n}$, Figure \ref{fig:geo}. The angular distribution for a polarized $\Lambda$ is given by: 

\begin{equation}
    \frac{dN}{d\cos\theta^*} = \frac{N}{2}(1+\alpha_\Lambda P\cos\theta^*)
    \label{eq:pol_fit}
\end{equation}

where $\alpha_{\Lambda} = 0.732 \pm 0.014$ is the parity-violating decay asymmetry for $\Lambda$ \cite{pdg}, and assuming CP conservation then $\alpha_\Lambda = -\alpha_{\bar{\Lambda}}$. The polarization value can then be extracted by fitting the $\cos\theta^*$ distribution in $x_F$ and $p_T$ bins, where $x_F$ is defined as: 
\begin{equation}
    x_F = \frac{p_z^*}{|max(p_z^*)|} = \frac{2p_z^*}{\sqrt{s_{NN}}} 
\end{equation}
the longitudinal momentum in the center-of-mass frame being $p_z^* = \sqrt{m^2+p_T^2}\sinh(y^*)$.

\section{Large Hadron Collider beauty Experiment}
The Large Hadron Collider beauty (LHCb) detector is a forward spectrometer with precision tracking, particle identification, and excellent vertex reconstruction \cite{lhcb}. The LHCb detector covers the forward region with pseudorapidity of $2 < \eta < 5$ and angular acceptance of  10 mrad - 250 (300) mrad in the vertical (horizontal) plane. The Vertex Locator (VELO) is a fundamental component that measures the trajectories of particles near the collision, which is ideal to reconstruct and distinguish the primary and secondary vertices. The particle identification (PID) components in the LHCb detector are important to reconstruct and separate hadronic decays properly and is optimal for muons, protons, kaons, and pions. \\
\subsection{Collider data: $pp$, $p$Pb, Pb$p$}
There are various data sets already available to be analyzed at LHCb to measure hyperon polarization in different collisions and energies, which includes $pp$ data with $\sqrt{s} = $ 5.02, 8.16, and 13 TeV and $p$Pb data with $\sqrt{s_{NN}} = $ 5.02 and 8.16 TeV.  The $p$Pb data is taken in two different collision configurations with different rapidity coverage: the forward configuration ($p$Pb, with the proton beam in the direction of the detector) $1.5 < y^* < 4$, and the backward configuration (Pb$p$, lead beam in the direction of the detector) $-5 < y^* < -2.5$. 
\subsection{Fixed-target data: SMOG}
Additionally, unique to LHCb is the ability to collect fixed-target data. The System for Measuring the Overlap with Gas (SMOG) was originally for luminosity calibration of colliding proton beams, accomplished through the injection of noble gas (He, Ar, and Ne) into the VELO while one of the circulating beams produces beam-gas collisions \cite{smog}. Since 2015, various data sets have been already collected with proton beam energies of 2500, 4000, and 6500 GeV. With SMOG, we can access central and backward rapidity, allowing for observation of particles with larger, negative $x_F$. 

\begin{figure}
    \centering
\includegraphics[width = 0.8\textwidth]{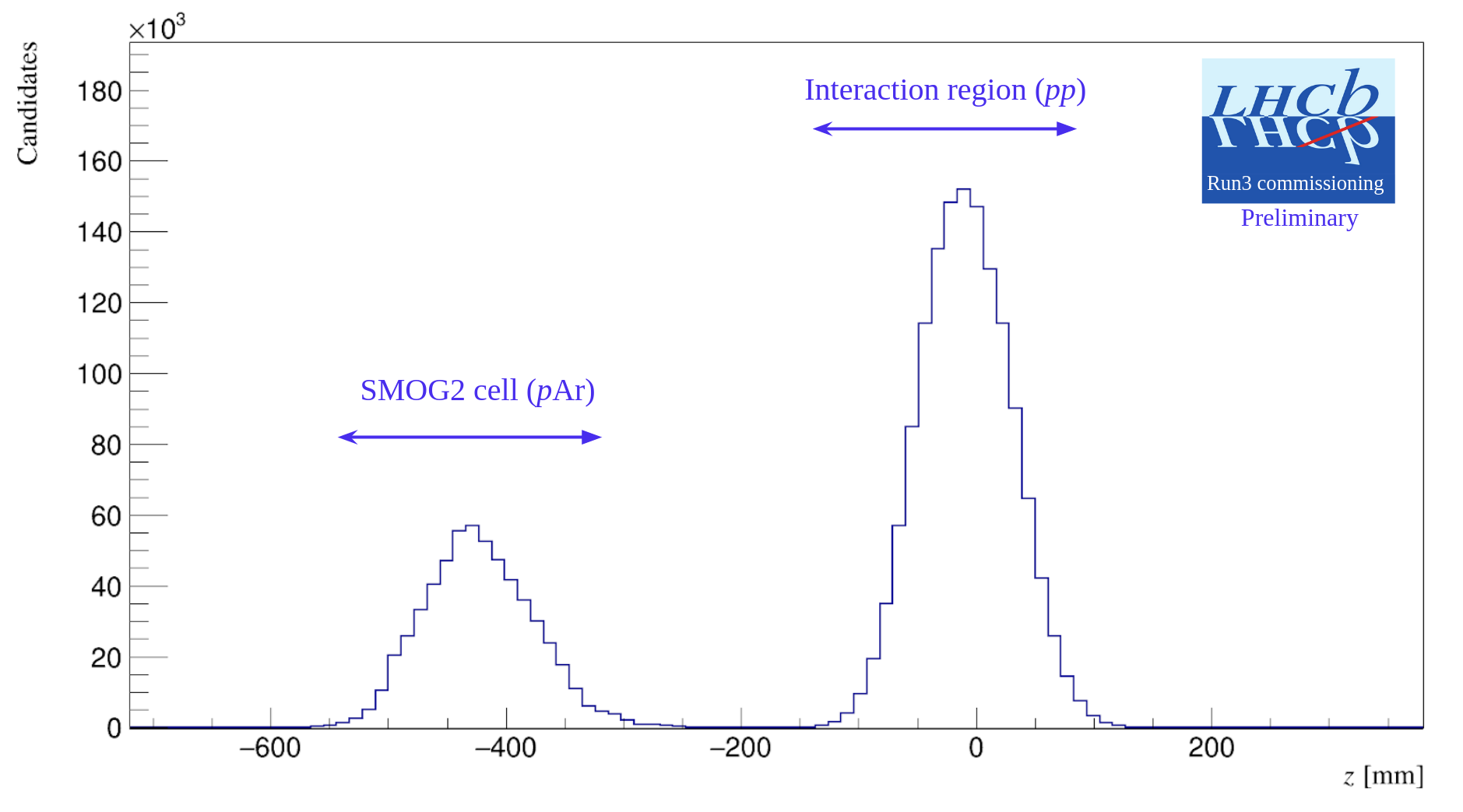}
    \caption{The longitudinal primary vertex distribution for data taken during 2022, where two clear separated regions for $pp$ and $p$Ar collisions can be observed. \cite{smog2}}
    \label{fig:smog}
\end{figure}

Between 2018 and 2022, the SMOG system was upgraded with the installation of a 20-cm-long confinement cell for the gas and a new injection system. The smaller beam-gas interaction region allows for higher injectable pressure and gas species, including heavy noble gases and non-noble species (He, Ne, Ar, Kr, Ze, H$_2$, D$_2$, N$_2$, O$_2$). Its separation with respect to the $pp$ was already demonstrated with 2022 LHC collision data, as exemplified by the distribution of reconstructed collision vertex $z$ coordinate as shown in Figure \ref{fig:smog}. 

\begin{figure}
    \centering
\includegraphics[width = 0.49\textwidth]{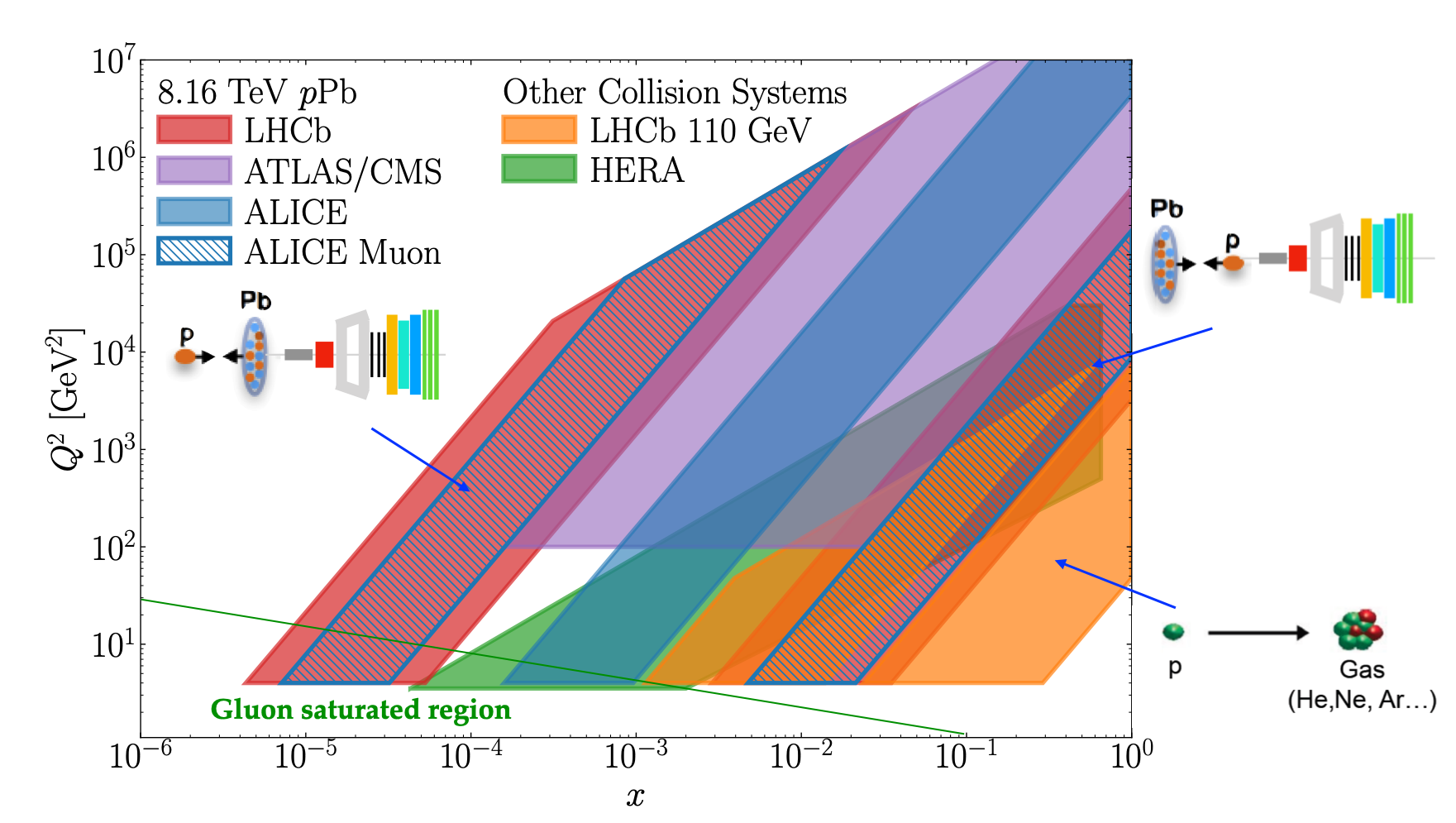}
\includegraphics[width= 0.49\textwidth]{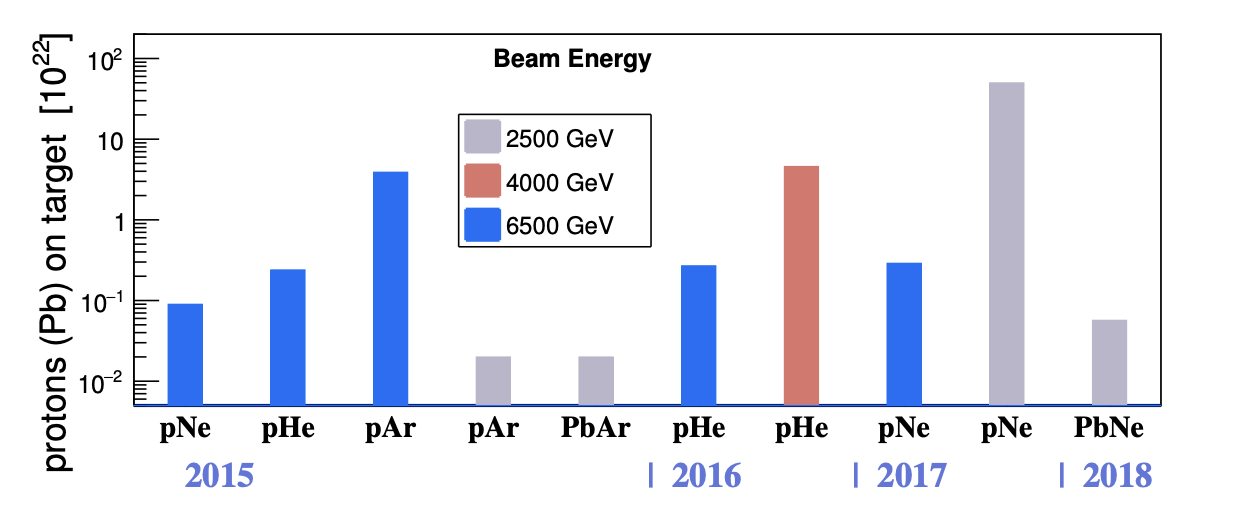}
    \caption{Left: The $x,Q^2$ kinematic coverage for LHCb heavy-ion in the forward ($p$Pb, red), backward (Pb$p$, red), and fixed-target (orange) configuration in comparison with other experiments. Right: SMOG data samples collected between 2015-2018.}
    \label{fig:kin}
\end{figure}

\section{Prospect of Spin Physics at LHCb}
\subsection{Transverse Hyperon Polarization}
The LHC beams high energy and the LHCb forward acceptance will be interesting to study transverse hyperon polarization at different energies and collision configurations as a function of $p_T$ and $x_F$ in a kinematic area that has been poorly explored. Additionally, the prospect for other transverse hyperon polarization measurements at LHCb has also been discussed; particularly of interest is analyzing the transverse polarization of $\Xi^-\rightarrow\Lambda\pi^-$ and $\Omega^-\rightarrow \Lambda K^-$. The measurement from the LHCb, along with $e^+e^-$ and SIDIS measurements, can put us in a better position to understand transverse hyperon polarization. 

\subsection{LHCSpin}

The LHCSpin project will also open up the capabilities of future spin and polarized physics measurement capabilities at LHCb \cite{lhcspin}. LHCSpin is not an approved LHCb project, but the R\&D is officially supported to add a transversely polarized target in the target storage cell by 2029. This will give insight on quark and gluon distributions at high $x$ and intermediate $Q^2$, and complementary measurements to existing and future SIDIS.


\begin{thebibliography}{12}

\bibitem{1976}
G. Bunce et al.,.
\newblock {$\Lambda^0$ Hyperon Polarization in Inclusive Production by 300-GeV
  Protons on Beryllium}.
\newblock {\em Phys. Rev. Lett.}, 36, 1113, 1976.

\bibitem{kpr}
G.L. Kane, J. Pumplin, W. Repko.
\newblock {Transverse Quark Polarization in Large $p_{T}$ Reactions,
  $e^{+}e^{-}$ Jets, and Leptoproduction: A Test of Quantum Chromodynamics}.
\newblock {\em Phys. Rev. Lett.}, 41, 1689, 1978.

\bibitem{belle}
Belle Collaboration.
\newblock {Observation of Transverse $\Lambda$ and $\bar{\Lambda}$ Hyperon
  Polarization in $e^+e^-$ Annihilation at Belle}.
\newblock {\em Phys. Rev. Lett.}, 122, 042001, 2019.

\bibitem{spin1996}
K.Heller.
\newblock {Spin and High Energy Hyperon Production Results and Prospects}.
\newblock {\em Proceedings of the 12th International Symposium on Spin
  Physics}, 1997.

\bibitem{twist}
Y. Koike, K.Yabe, and S. Yoshida.
\newblock {Hyperon polarization from the twist-3 distribution in unpolarized
  proton-proton collision}.
\newblock {\em Phys. Rev. D}, 92, 094011, 2015.

\bibitem{pff}
U. D'Alesio, F. Murgia, M. Zaccheddu.
\newblock {First extraction of the $\Lambda$ polarizing fragmentation function
  from Belle $e^+e^-$ data}.
\newblock {\em Phys. Rev. D.}, 102, 054001, 2020.

\bibitem{elliot}
Elliot Leader.
\newblock {\em {Spin in Particle Physics}}.
\newblock Cambridge University Press, 2001.

\bibitem{pdg}
P.A. Zyla et al (Particle Data Group).
\newblock {Prog. Theor. Exp. Phys.}
\newblock {\em Phys. Rev. D.}, 083C01, 2020.

\bibitem{lhcb}
The LHCb Collaboration.
\newblock {The LHCb Detector at the LHC}.
\newblock {\em JINST}, S0800, 2008.

\bibitem{smog}
The LHCb Collaboration.
\newblock {LHCb SMOG Upgrade}.
\newblock {LHCB-TDR-020}.
\newblock Technical report, 2019.

\bibitem{smog2}
The LHCb Collaboration.
\newblock {First LHCb upgrade reconstruction results on fixed-target data}.
\newblock 2023.
\newblock LHCb-FIGURE-2023-001.

\bibitem{lhcspin}
C. A. Aidala, et al.
\newblock {The LHCSpin Project}.
\newblock 2018.
\newblock arXiv:1901.08002.

\end{thebibliography}
\end{document}